\documentclass{article}

\usepackage{amsfonts}
\usepackage[russian,english]{babel}
\usepackage{booktabs}
\usepackage{dirtree}
\usepackage{fontawesome}
\usepackage[T2A,T1]{fontenc}
\usepackage{hyperref}
\usepackage[utf8]{inputenc}
\usepackage{listings}
\usepackage{microtype}
\usepackage[numbers]{natbib}
\usepackage[preprint, nonatbib]{neurips} 
\usepackage{nicefrac}
\usepackage{pgf-pie}
\usepackage{pgfplots}
\pgfplotsset{width=10cm,compat=1.9}
\usepackage{xcolor}
\usepackage{xspace}
\usepackage{subcaption}
\usepackage{tcolorbox}
\usepackage{url}

\tcbuselibrary{skins}
\usepackage[acronym]{glossaries}
\newacronym{asr}{ASR}{Automatic Speech Recognition}
\newacronym{cer}{CER}{Character Error Rate}
\newacronym{mad}{MAD}{Movie Audio Descriptions}
\newacronym{mlb}{MLB}{Major League Baseball}
\newacronym{nlp}{NLP}{Natural Language Processing}
\newacronym{vad}{VAD}{Voice Activity Detection}
\newacronym{wer}{WER}{Word Error Rate}

\newcommand{\mysection}[1]{\vspace{2pt}\noindent\textbf{#1}}
\newcommand{\dsname}[0]{SoccerNet-Echoes\xspace}
\newcommand{\dsurl}[0]{\url{https://github.com/SoccerNet/sn-echoes}\xspace}

\title{\dsname: A Soccer Game Audio Commentary Dataset}

\author{Sushant Gautam$^{1,2}$ 
\quad Mehdi Houshmand Sarkhoosh$^{2,3}$ 
\quad Jan Held$^{5}$ 
\quad Cise Midoglu$^{1,3}$ \\
\textbf{Anthony Cioppa}$^{5,6}$ 
\quad \textbf{Silvio Giancola}$^{6}$ 
\quad \textbf{Vajira Thambawita}$^{1}$  \\
\quad \textbf{Michael A. Riegler}$^{1}$ 
\quad \textbf{Pål Halvorsen}$^{1,2,3}$ 
\quad \textbf{Mubarak Shah}$^{4}$\\
$^1$\textit{SimulaMet, Norway} \quad 
$^2$\textit{OsloMet, Norway} \quad 
$^3$\textit{Forzasys, Norway} \\
$^4$\textit{University of Central Florida, USA}  \quad 
$^5$\textit{University of Liège, Belgium} \quad $^6$\textit{KAUST, Saudi Arabia}\\
\dsurl
}

\begin{document}
\maketitle

\begin{abstract}

The application of \gls{asr} technology in soccer offers numerous opportunities for sports analytics. Specifically, extracting audio commentaries with \gls{asr} provides valuable insights into the events of the game, and opens the door to several downstream applications such as automatic highlight generation. This paper presents \dsname, an augmentation of the SoccerNet dataset with automatically generated transcriptions of audio commentaries from soccer game broadcasts, enhancing video content with rich layers of textual information derived from the game audio using \gls{asr}. These textual commentaries, generated using the \emph{Whisper} model and translated with \emph{Google Translate}, extend the usefulness of the SoccerNet dataset in diverse applications such as enhanced action spotting, automatic caption generation, and game summarization. By incorporating textual data alongside visual and auditory content, \dsname aims to serve as a comprehensive resource for the development of  algorithms specialized in capturing the dynamics of soccer games. We detail the methods involved in the curation of this dataset and the integration of \gls{asr}. We also highlight the implications of a multimodal approach in sports analytics, and how the enriched dataset can support diverse applications, thus broadening the scope of research and development in the field of sports analytics.

\end{abstract}

\section{Introduction}\label{sec:introduction}

\begin{figure*}[ht!]
    \centering
    \includegraphics[width=0.95\textwidth]{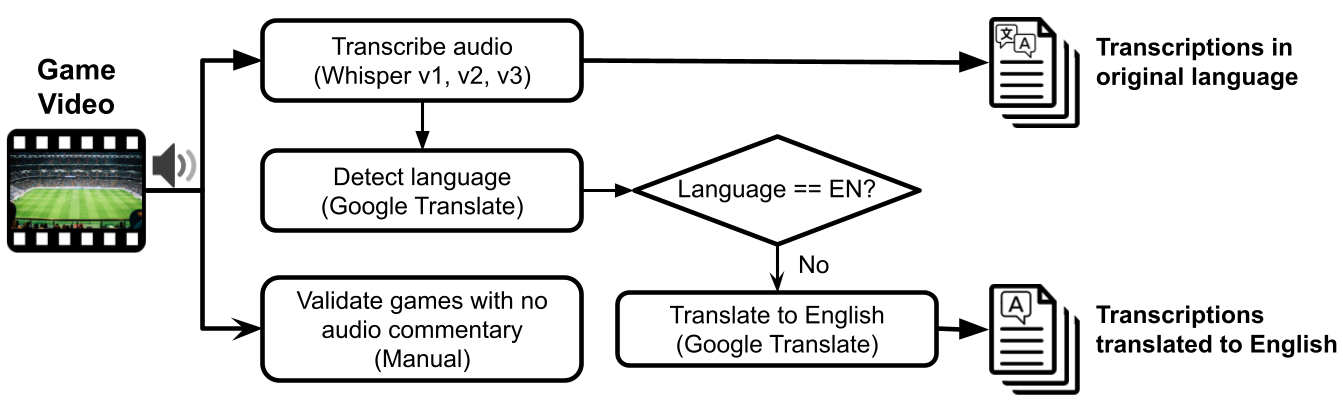}
    \caption{The pipeline for generating a multilingual commentary dataset from soccer game videos, incorporating different Whisper versions, language detection and translation.}
    \label{fig:pipeline}
\end{figure*}

Sports analytics has progressively embraced technological innovations, to enrich the analysis and understanding of complex sports events. Historically, analysis of sports video footage relied primarily on visual cues: action spotting, player recognition and player tracking were based solely on videos and video frames as images. However, with the advancement in multimodal data integration, the scope of sports analytics has broadened, encompassing audio and textual data alongside traditional visual input. This integration facilitates a more holistic view of sports events, capturing the nuances of game dynamics that visual data alone might miss.

The SoccerNet dataset~\cite{SoccerNet, SoccerNetv2}, initially designed to support video-based analyses such as action spotting, player tracking~\cite{Cioppa2022SoccerNetTracking}, camera calibration~\cite{Cioppa2022Scaling}, dense video captioning~\cite{Mkhallati2023SoccerNetCaption}, foul recognition~\cite{Held2023VARS, Held2024XVARS}, and game state reconstruction~\cite{Somers2024SoccerNetGameState}, has emerged as a key resource in sports video analytics. Yet, the rapidly evolving field of sports analytics demands ever more sophisticated tools that can process and interpret the complex interplay of multiple data modalities, including video, audio, and text. In response to this need, we augment the SoccerNet dataset by incorporating state-of-the-art \gls{asr} technology to enrich the dataset with textual information derived from live game commentaries. This multimodal approach aims to refine the accuracy of existing tasks such as action spotting and dense video captioning, as well as extends the dataset's usefulness in more complex applications such as sentiment analysis or tactical assessment. 

This paper details our augmentation of the SoccerNet dataset to include audio commentary transcriptions, generated with OpenAI’s Whisper~\cite{WhisperGitHub} tool, and translated with Google Translate~\cite{GoogleTranslate}. The resulting \dsname dataset serves as a comprehensive toolkit for researchers and practitioners, allowing enhanced interpretation and narrative generation from sports footage. Through the integration of audio commentaries in textual format, \dsname facilitates a deeper understanding of the contextual aspects of soccer games, thereby enhancing both the analytical capabilities and the viewer's experience.

\mysection{Contributions.} Our contributions are threefold: (i) We augment the SoccerNet dataset with automatically transcribed and translated audio commentaries from soccer game broadcasts in multiple languages, and present this new dataset (\dsname) as an open-access resource. (ii) We evaluate \dsname using human-verified \gls{asr} from literature and perform quantitative analysis. (iii) We explore the potential applications of \dsname in various research areas, and highlight the broader implications of adopting a multimodal approach for sports research in general. 

\section{Background and Related Work}\label{section:background}

The integration of \gls{asr} technology has significantly advanced the analysis and processing of video content, impacting sports research~\cite{Malik2021Mar, Li2022Apr, Lu2020Sep, Chang1996}. Initially, video understanding focused primarily on computer vision tasks, such as action recognition and event spotting~\cite{Soomro2012Dec, Szeliski2022, Shih2017Jan, cabado2024beyond}. Advancements in dense video captioning~\cite{Krishna} began to explore the generation of captions for temporally localized activities within untrimmed videos, representing a significant departure from the traditional approach of generating a single caption for short clips. Further evolution in the field saw the incorporation of audio and speech modalities alongside visual data, enhancing video captioning capabilities.

\subsection{Early Applications: Action Recognition and Event Detection}

Before the integration of \gls{asr}, the primary focus in sports video analysis was on visual cues for game action recognition and event detection (spotting)~\cite{SoccerNet, SoccerNetv2}. Researchers began to explore the use of multimodal sports data, combining game audio and video streams to improve the accuracy of action spotting in soccer videos~\cite{Barnard2003, Vanderplaetse2020}. These studies marked an early recognition of the potential benefits of incorporating audio data alongside traditional video processing, setting the stage for the subsequent integration of \gls{asr} technologies in sports game analysis~\cite{Gautam2023Oct}.

\subsection{Transition to Speech and Language Processing}

As the potential of multimodal analysis became evident, subsequent studies began to integrate more sophisticated speech and language processing technologies~\cite{Gautam2022Oct, gautam2022assisting}. Gao et al.~\cite{Gao2020} utilized \gls{asr} to segment and categorize commentary from soccer videos, thereby enhancing key moment extraction and highlight generation. This application of \gls{asr} for segmenting commentator speech represented a shift towards leveraging linguistic information to complement visual data. Gautam et al. used captions, text and commentaries extracted from \gls{asr} for game summarization~\cite{Gautam2022Oct}, and used audio intensity to capture the field excitement~\cite{gautam2022assisting}.

\subsection{Multimodal Dense Video Captioning}

The application of textual modality in sports video analysis reached a new level with the introduction of dense video captioning tasks, as exemplified by the SoccerNet Challenge~\cite{Cioppa2023Sep}. The SoccerNet-Caption dataset~\cite{Mkhallati2023} leverages rich, timestamped textual commentaries that capture both factual and emotional aspects of the game. Aiming to bridge the gap for fans unable to watch soccer games live by providing engaging summaries that mimic the excitement of real-time commentary, this task demonstrated the growing importance of textual modality in creating immersive fan experiences. 

Lashin et al.~\cite{Lashin2020Jun} highlighted the transformative potential of integrating \gls{asr} alongside video and audio data in dense video captioning. Their multimodal approach significantly improved event description accuracy, emphasizing the vital contextual cues provided by \gls{asr} that often elude video-only analyses. By combining visual information with speech data, their method demonstrated substantial performance enhancements, showcasing the synergistic benefits of leveraging multimodal inputs for comprehensive video understanding. \gls{mad} dataset offers a novel benchmark for video-language grounding tasks, by aligning audio descriptions of mainstream movies with video content~\cite{Soldan_2022_CVPR}. The multimodal approach has since been embraced~\cite{Seo2022Jun}, enriching the scope of video captioning and commentary generation~\cite{GOALQi2023Oct, Sattar2023cricket}, and illustrating the transformative impact of \gls{asr} technology in video content analysis and processing.

\subsection{Advancements in ASR Technologies}

The development of LLM-based \gls{asr} technologies~\cite{WhisperPaper} further accelerated the use of speech recognition in sports analytics. The GOAL dataset~\cite{GOALQi2023Oct}, derived from SoccerNet videos, comprises 80 full-game videos transcribed into raw text using the Azure ASR toolkit, then meticulously curated by English-speaking, soccer-knowledgeable annotators. 20 games were selected after quality filtering, and annotation tasks included proofreading, aligning text with video timelines, and annotating knowledge entities, ensuring comprehensive linguistic and visual analysis. Ikeda et al.~\cite{Ikeda2024Feb} leveraged \gls{asr} alongside Whisper~\cite{WhisperPaper} and a speech recognition model trained on the extensive Japanese audio corpus ReazonSpeech~\cite{yin2023reazonspeech} to transcribe commentary from 134 Japanese \gls{mlb} highlight videos. To address noise challenges, Whisper transcripts underwent manual labeling, while ReazonSpeech transcripts were labeled using keyword techniques, thereby improving dataset quality and reducing time overheads. These approaches underscore a growing trend in leveraging \gls{asr} technology to create rich datasets for detailed examination of both linguistic and visual elements in sports commentary.

\subsection{Broader Applications and Future Directions}

The integration of \gls{asr} technology into soccer video analysis has followed a trajectory from enhancing basic action recognition to enabling sophisticated, multimodal interpretations of complex events. The evolution of this technology has not only improved the analytical capabilities of researchers but has also significantly enhanced the viewing experience for fans worldwide. Beyond soccer, the integration of \gls{asr} has also influenced broader video analysis tasks. Hessel et al.~\cite{Hessel2019Nov} explored combining \gls{asr} and visual features to generate captions for instructional videos, highlighting how \gls{asr} can aid in distinguishing between similar actions through linguistic cues. Sattar et al.~\cite{Sattar2023cricket} focused on identifying events in cricket games by combining commentary text obtained from \gls{asr} with visual data and cues such as replays, bowler and umpire positions. 

These applications suggest potential future directions where \gls{asr} could be integrated into various types of content beyond sports, enriching content accessibility and understanding across domains. As \gls{asr} technology continues to advance, its applications in video analysis are expected to expand, further revolutionizing the field and providing deeper insights into the intricate dynamics of media content. However, it is important to acknowledge that despite the advancements, \gls{asr} technology is not flawless. Automated capture introduces uncertainty and errors in the ground truth data. Nonetheless, its cost-effectiveness and scalability make \gls{asr} a promising candidate for widespread use.

\section{Dataset Curation}\label{sec:dataset}

Our comprehensive approach to curating a multilingual soccer commentary dataset, which includes the automated transcription and translation of commentaries, is illustrated in Figure~\ref{fig:pipeline}. Below, we detail the curation of the \dsname using the illustrated methodology.

\subsection{Sources and Collection}

\begin{figure}[htbp]

    \centering
    
    \pgfplotsset{width=\columnwidth, compat=1.9}
    
    \begin{tikzpicture}
    
        \begin{axis}[
            ybar,
            height=0.3\textheight,
            symbolic x coords={English, Spanish, Russian, German, French, Turkish, Italian, Polish, Bosnian, Hungarian, Not Available},
            xtick=data,
            nodes near coords,
            nodes near coords align={vertical},
            ylabel={Count},
            x tick label style={rotate=45, anchor=east},
            ymin=0,
            axis x line*=bottom,
            axis y line*=left,
            ]
            \addplot coordinates {
              (English, 297)
                (Spanish, 264)
                (Russian, 218)
                (German, 135)
                (French, 102)
                (Turkish, 4)
                (Italian, 4)
                (Polish, 2)
                (Bosnian, 2)
                (Hungarian, 2) 
                (Not Available, 70)
            };
        \end{axis}
        
    \end{tikzpicture}

    \caption{\textbf{Language distribution} of the original broadcast audio for each game half, identified using Google Translate and the Whisper language detection model (1100 in total).}
    \label{fig:dataset-original-language}
    
\end{figure}
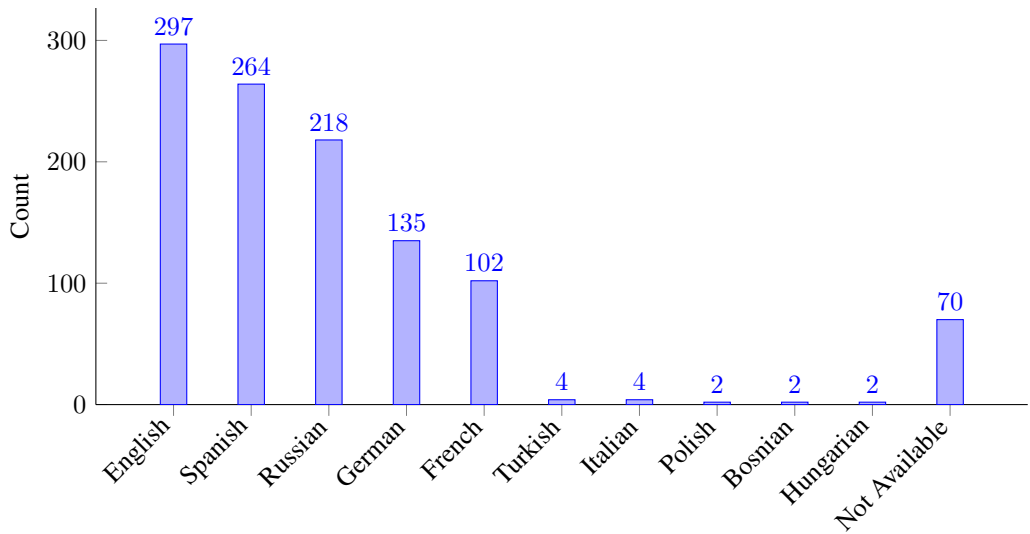

We used all 1100 game half videos from the SoccerNet~\cite{SoccerNet} dataset. These correspond to 550 games from 6 different leagues and 4 different seasons, as depicted in Figure~\ref{fig:dataset-count-per-league}. The games were narrated live by commentators speaking one of 10 different languages. The distribution of the original language of the broadcast game audio is provided in Figure~\ref{fig:dataset-original-language}. We automatically extracted commentaries in textual format following the procedure described below. 

\begin{figure}[htbp]

    \centering
    \begin{tikzpicture}
        \begin{axis}[
            width=1\columnwidth,
            height=0.3\textheight,
            ybar=10pt, 
            bar width=10pt, 
            legend style={at={(0.03,0.97)}, anchor=north west, cells={align=left}}, 
            ylabel={Number of Games}, 
            symbolic x coords={EPL, UCL, Ligue1, Bundesliga, SerieA, LaLiga}, 
            xtick=data, 
            x tick label style={rotate=45,anchor=east}, 
            axis x line*=bottom, 
            axis y line*=left, 
            ]
            
        \addplot+[bar shift=-12pt] coordinates {(EPL,6) (Ligue1,1) (SerieA,11) (Bundesliga,8) (LaLiga,18) (UCL,37)}; 
        \addlegendentry{2014-2015} 
        
        \addplot+[bar shift=0pt] coordinates {(EPL,49) (UCL,45) (Ligue1,3) (Bundesliga,18) (SerieA,9) (LaLiga,36)}; 
        \addlegendentry{2015-2016} 
        
        \addplot+[bar shift=12pt] coordinates {(EPL,49) (UCL,26) (Ligue1,43) (Bundesliga,35) (SerieA,85) (LaLiga,63)}; 
        \addlegendentry{2016-2017} 
        
        \addplot+[bar shift=24pt] coordinates {(LaLiga,8)}; 
        \addlegendentry{2019-2020} 
        
        \end{axis}
    \end{tikzpicture}
    
    \caption{\textbf{Number of games} per league and season in the SoccerNet dataset (550 in total).}
    \label{fig:dataset-count-per-league}

\end{figure}
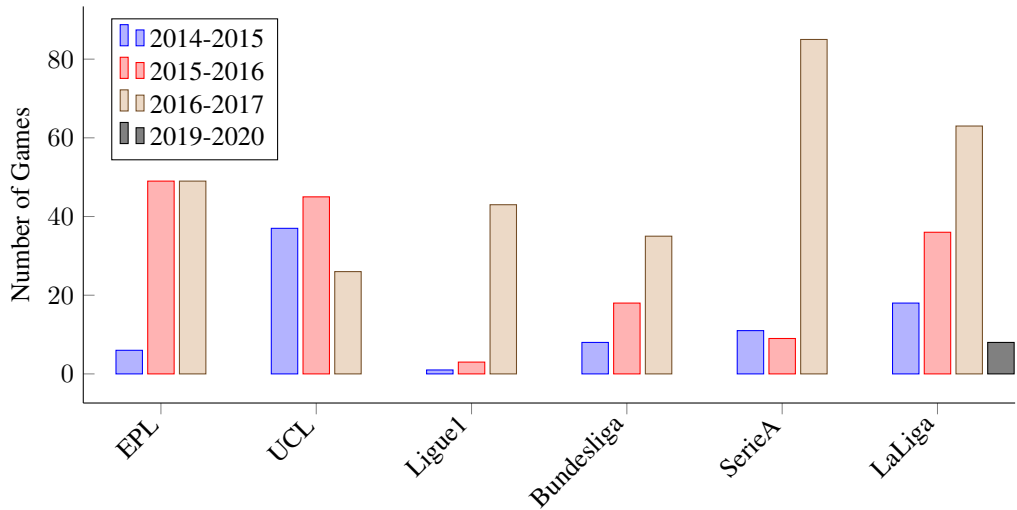

\subsection{Transcriptions}

\begin{figure*}[ht!]

    \centering
    \begin{subfigure}{0.19\textwidth}
        \includegraphics[width=\linewidth]{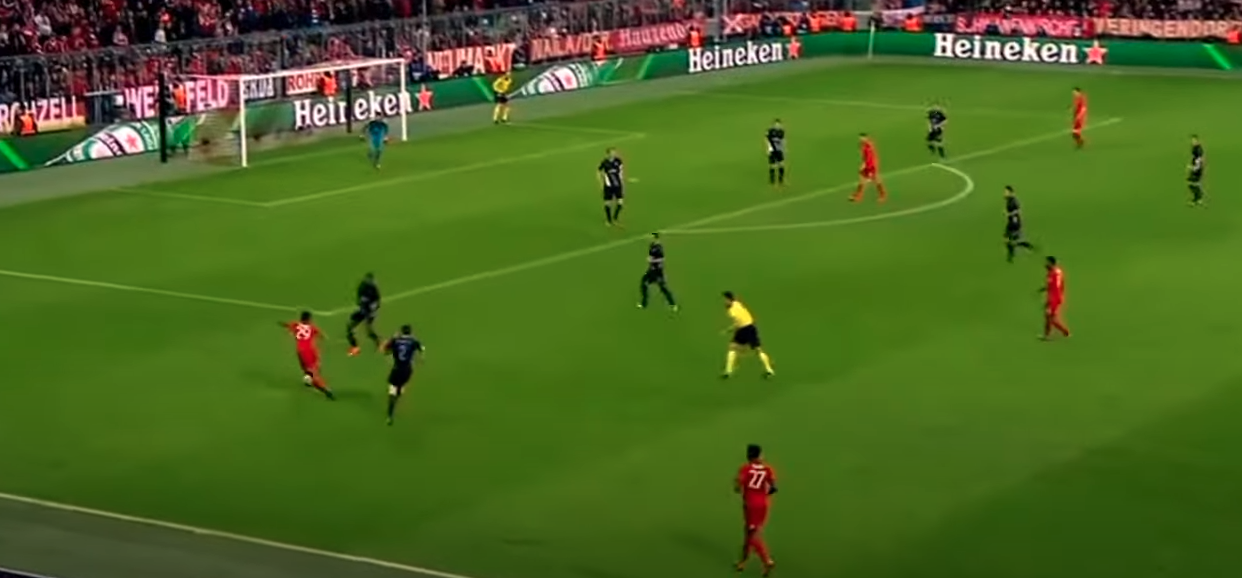}
        \caption{}
    \end{subfigure}
    \hfill
    \begin{subfigure}{0.19\textwidth}
        \includegraphics[width=\linewidth]{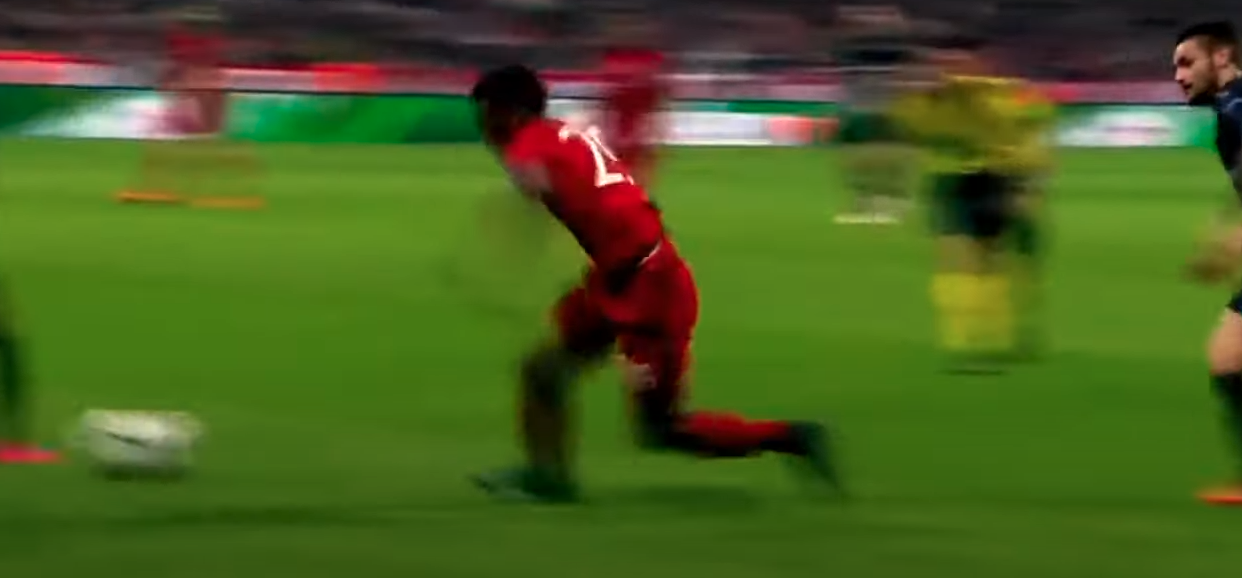}
        \caption{}
    \end{subfigure}
    \hfill
    \begin{subfigure}{0.19\textwidth}
        \includegraphics[width=\linewidth]{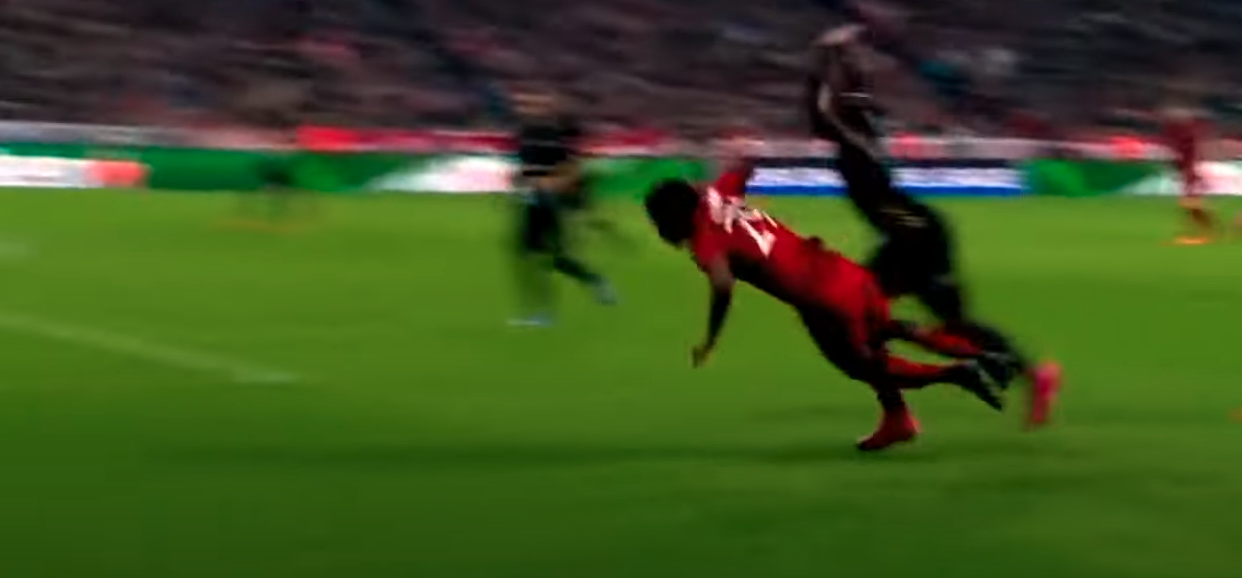}
        \caption{}
    \end{subfigure}
    \hfill
    \begin{subfigure}{0.19\textwidth}
        \includegraphics[width=\linewidth]{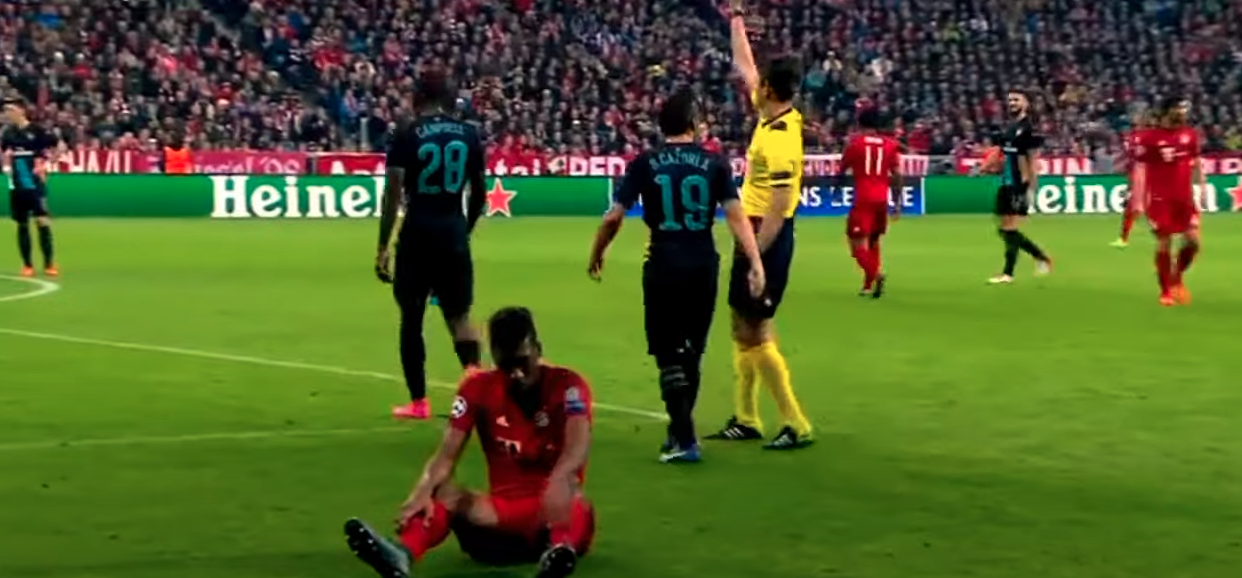}
        \caption{}
    \end{subfigure}
    \hfill
    \begin{subfigure}{0.19\textwidth}
        \includegraphics[width=\linewidth]{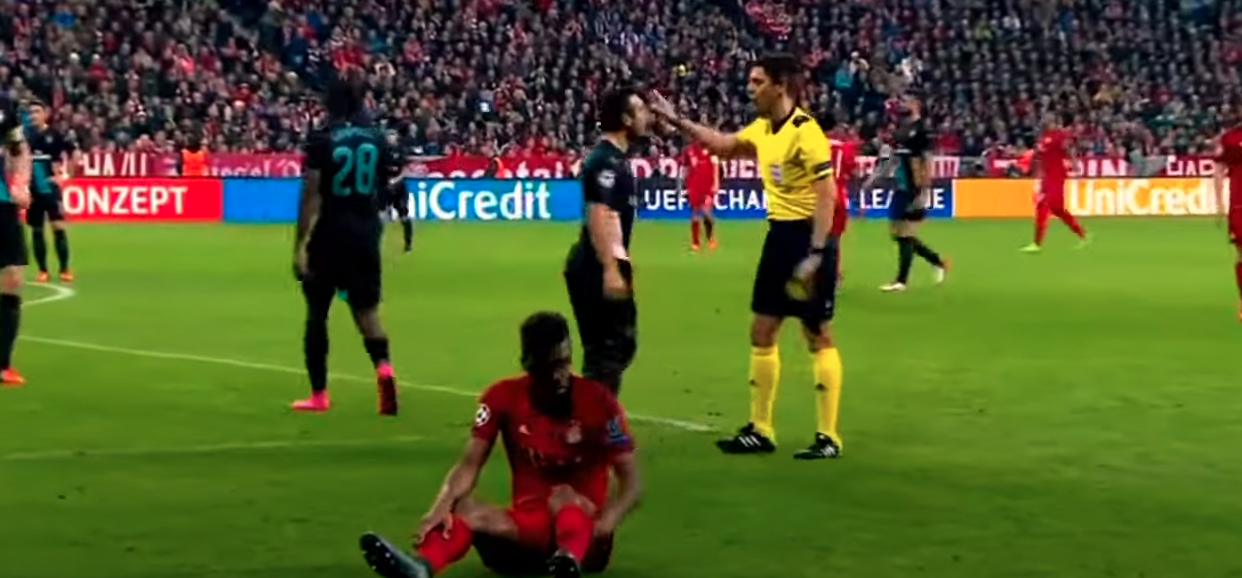}
        \caption{}
    \end{subfigure}
    \label{fig:asr-series}
    
    \begin{tcolorbox}[colback=white,colframe=blue!75!black,boxrule=0.5mm,arc=2mm,outer arc=2mm,boxsep=2mm,left=1mm,right=1mm,top=1mm,bottom=1mm, title=Automatic English Transcription]
    
        \textbf{Frame a to e:} "Here is Coman. I wonder if the referee's going to book him for that because that looked like a blatant dive really."\\
        
    \end{tcolorbox}
    
    \caption{Example transcription using Whisper (original audio in English). Key frames are shown to represent the corresponding video segment.}
    \label{fig:asr-example-en}

\end{figure*}

We used OpenAI's Whisper large-v1\footnote{\url{https://huggingface.co/openai/whisper-large}}, large-v2\footnote{\url{https://huggingface.co/openai/whisper-large-v2}}, and large-v3\footnote{\url{https://huggingface.co/openai/whisper-large-v3}} \gls{asr} models to transcribe the commentator speech from the game audio into text, with the default parameters provided in the original GitHub implementation (commit: \texttt{ba3f3cd})~\cite{WhisperGitHub}. Initially, the \gls{asr} models were used to process all games regardless of the presence of audible human commentary. By default, Whisper models recognize the audio language from the first $30$ seconds of the input, and assume that the rest of the audio is in the same language. It should be noted that the model's capability to accurately identify entity names (players, teams, stadiums, etc.) is limited. We performed explicit language detection for all games using Whisper-v2, the resulting language distribution is depicted in Figure \ref{fig:dataset-original-language}. Figure~\ref{fig:asr-example-en} presents an example transcription using Whisper from original game audio in English. 

\mysection{Manual validation of videos without commentary.} As shown in Figure~\ref{fig:dataset-original-language}, 70 game half videos in the SoccerNet dataset lack audio commentary and therefore had empty \gls{asr} outputs. We manually inspected these assets to ensure that they do not in fact contain commentary. Overall, 56 videos were confirmed not to have game audio at all, and 14 videos were confirmed to have game audio but without commentary (only stadium). The detailed list of videos in the SoccerNet dataset without audio commentary is released along with the \dsname dataset as a spreadsheet.

\subsection{Translations}

\begin{figure*}[h!]

    \centering
    \begin{subfigure}{0.19\textwidth}
        \includegraphics[width=\linewidth]{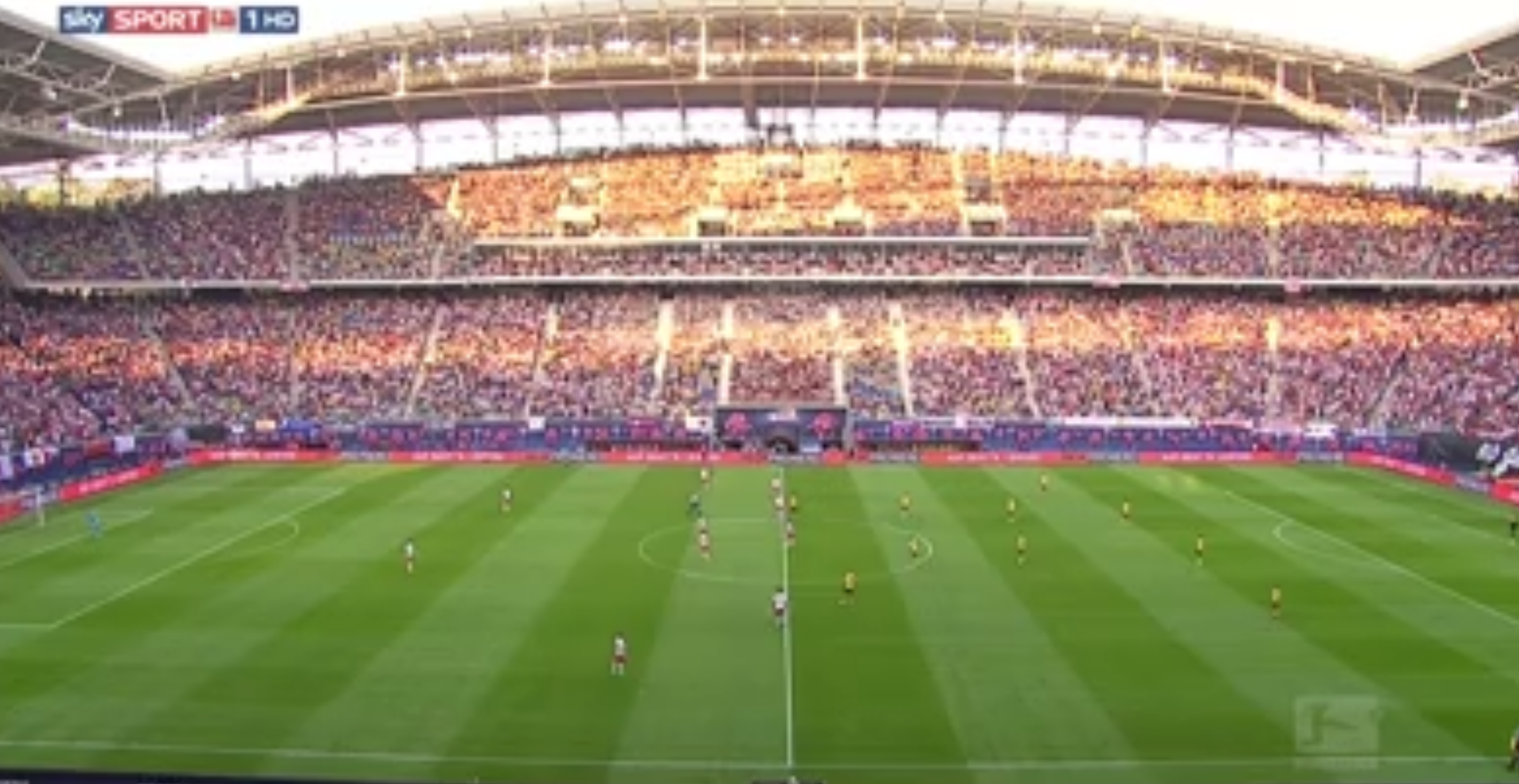}
        \caption{}
        \label{fig:asr-example-de-1}
    \end{subfigure}
    \hfill
    \begin{subfigure}{0.19\textwidth}
        \includegraphics[width=\linewidth]{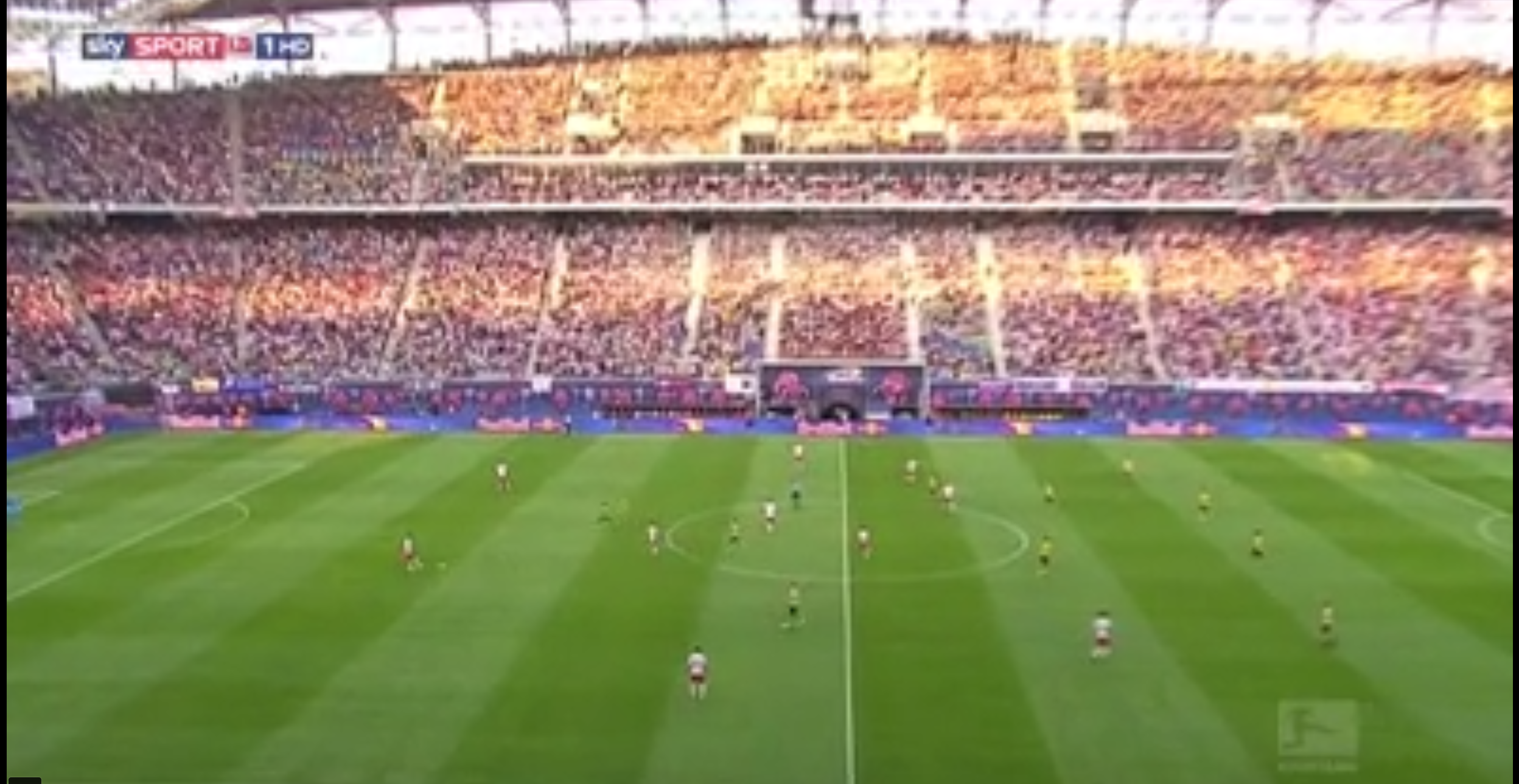}
        \caption{}
        \label{fig:asr-example-de-2}
    \end{subfigure}
    \hfill
    \begin{subfigure}{0.19\textwidth}
        \includegraphics[width=\linewidth]{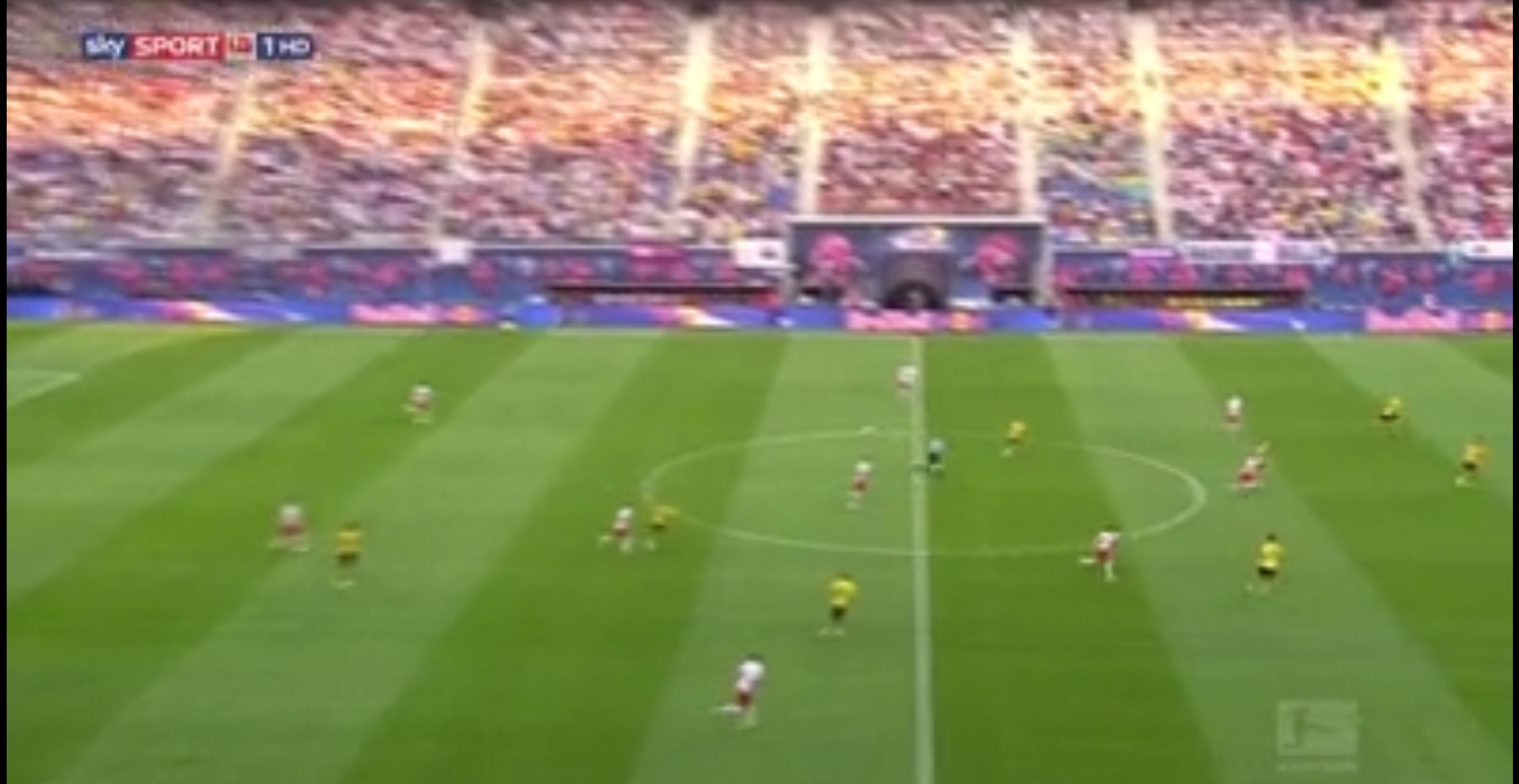}
        \caption{}
        \label{fig:asr-example-de-3}
    \end{subfigure}
    \hfill
    \begin{subfigure}{0.19\textwidth}
        \includegraphics[width=\linewidth]{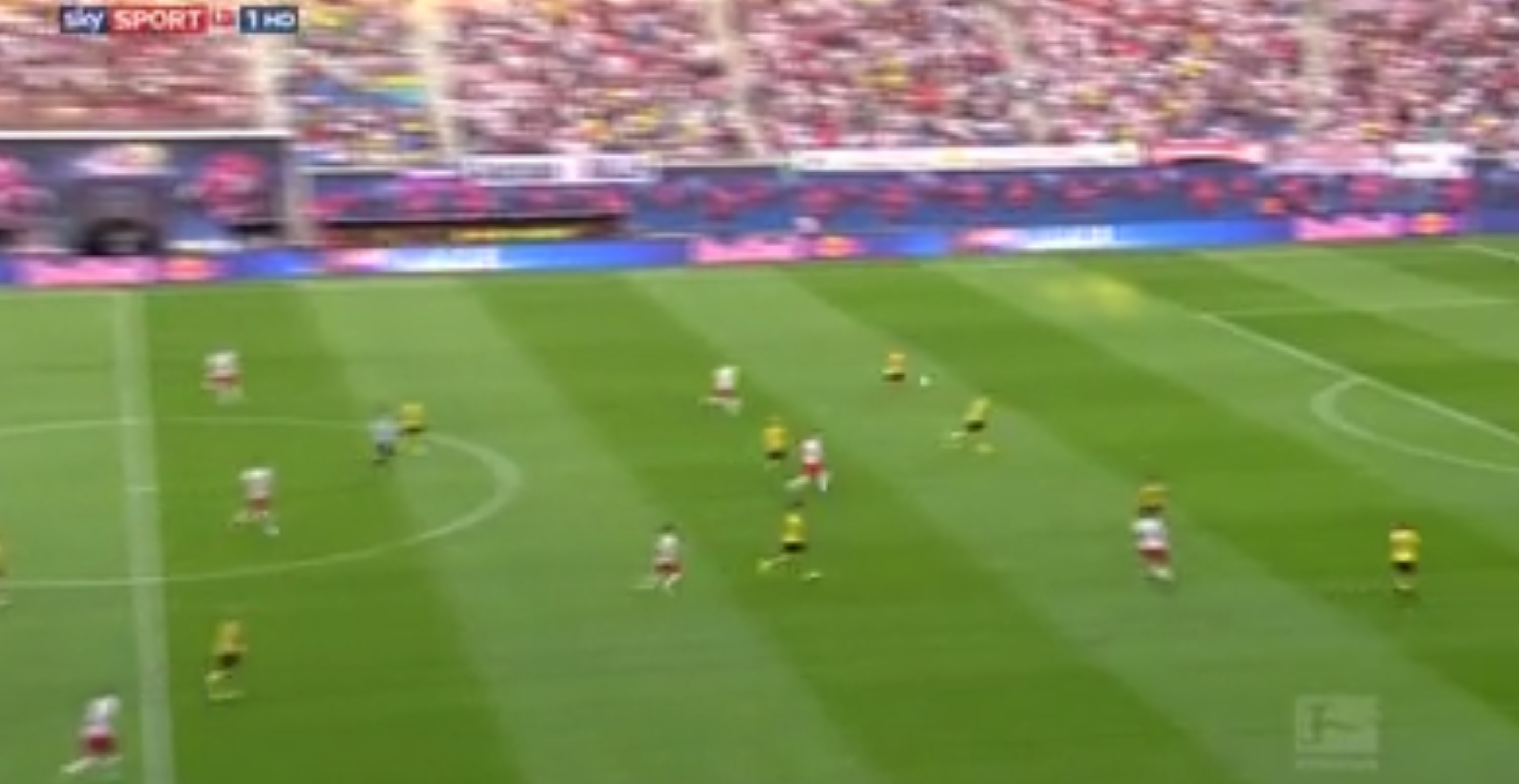}
        \caption{}
        \label{fig:asr-example-de-4}
    \end{subfigure}
    \hfill
    \begin{subfigure}{0.19\textwidth}
        \includegraphics[width=\linewidth]{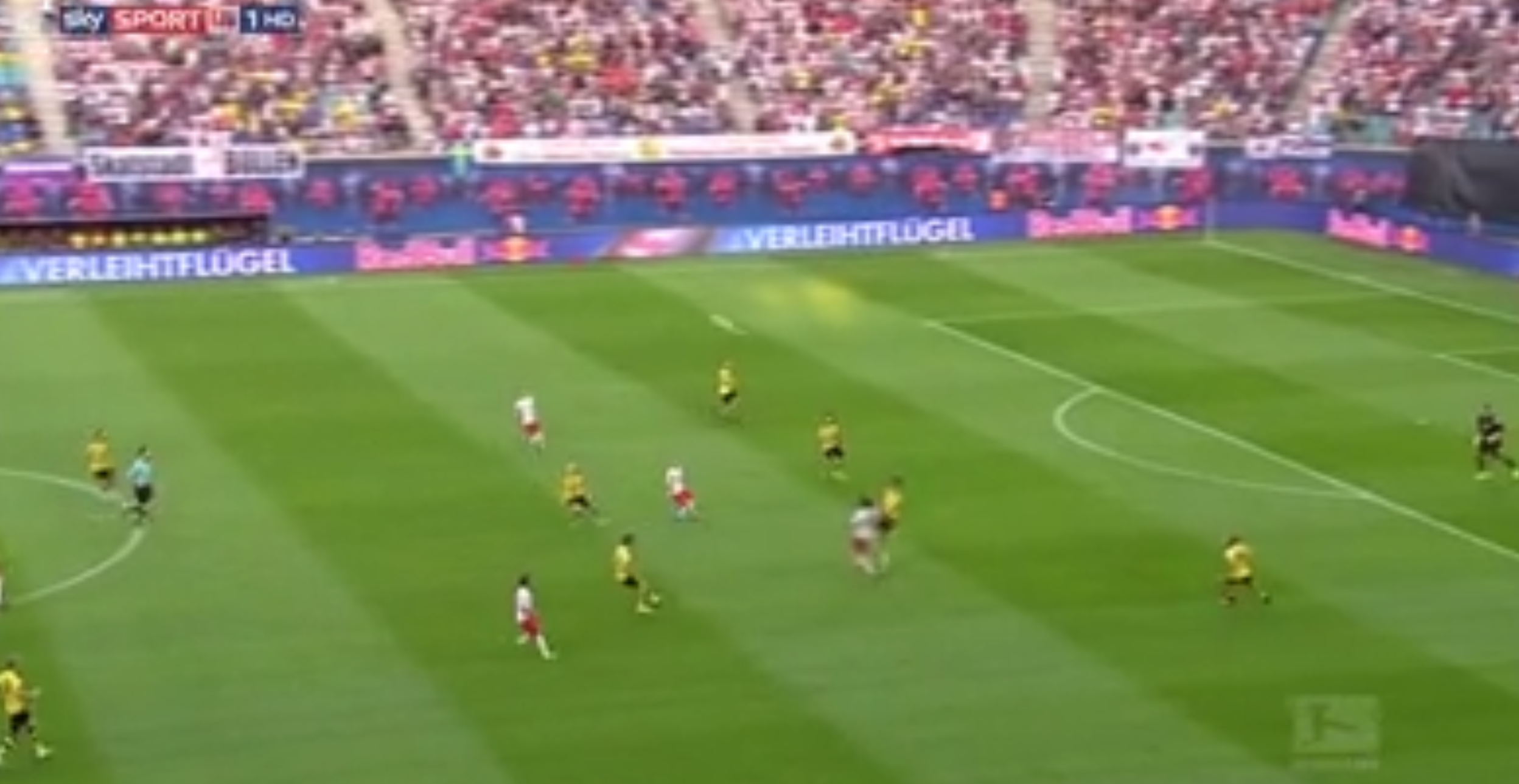}
        \caption{}
        \label{fig:asr-example-de-5}
    \end{subfigure}
    
    \begin{tcolorbox}[colback=white,colframe=blue!75!black,boxrule=0.5mm,arc=2mm,outer arc=2mm,boxsep=2mm,left=1mm,right=1mm,top=1mm,bottom=1mm, title=Automatic German Transcription and English Translation]
    
        \textbf{Frame a to e (Transcription):} "Spiel für RB Leipzig, die Mannschaft spielt in weiß und rot Borussia Dortmund im ersten"\\
        
        \textbf{Frame a to e (Translation):} "Game for RB Leipzig, the team plays in white and red Borussia Dortmund in the first"\\
    
    \end{tcolorbox}
    
    \caption{Example transcription using Whisper and translation to English using Google Translate (original audio in German). Key frames are shown to represent the corresponding video segment.}
    \label{fig:asr-example-de}

\end{figure*}

\begin{figure*}[ht!]

    \centering
    \begin{subfigure}{0.19\textwidth}
        \includegraphics[width=\linewidth]{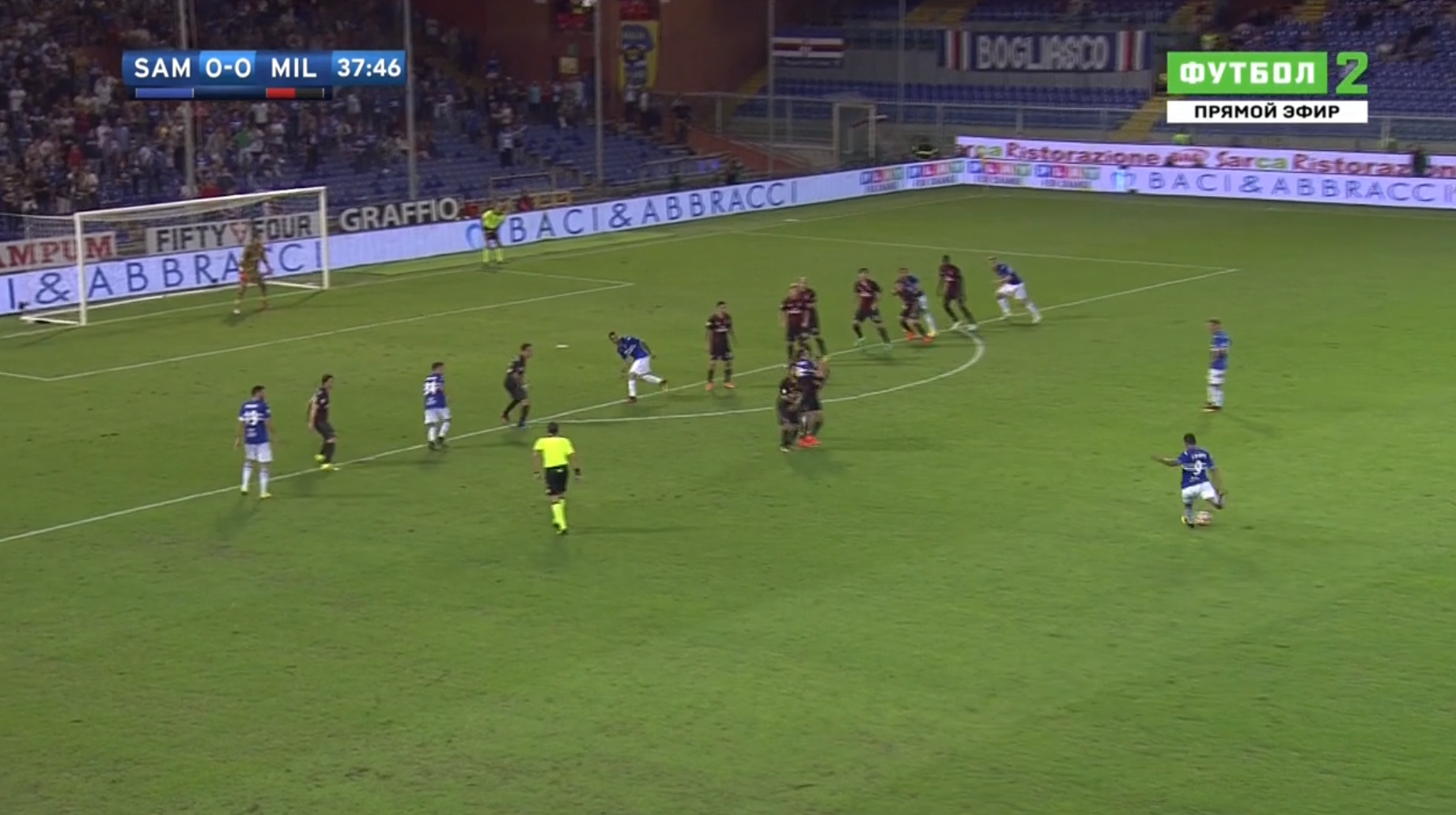}
        \caption{}
    \end{subfigure}
    \hfill
    \begin{subfigure}{0.19\textwidth}
        \includegraphics[width=\linewidth]{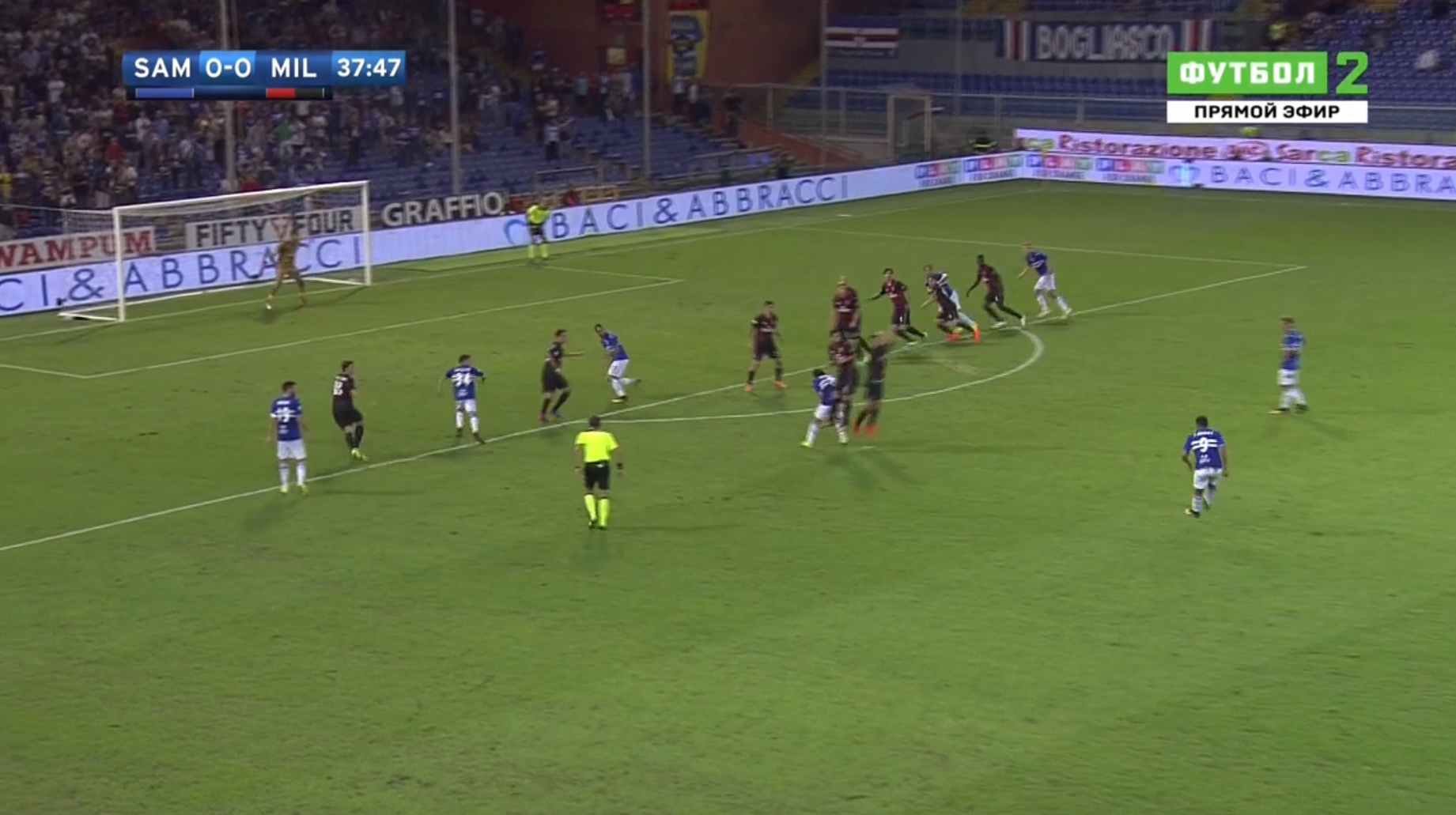}
        \caption{}
    \end{subfigure}
    \hfill
    \begin{subfigure}{0.19\textwidth}
        \includegraphics[width=\linewidth]{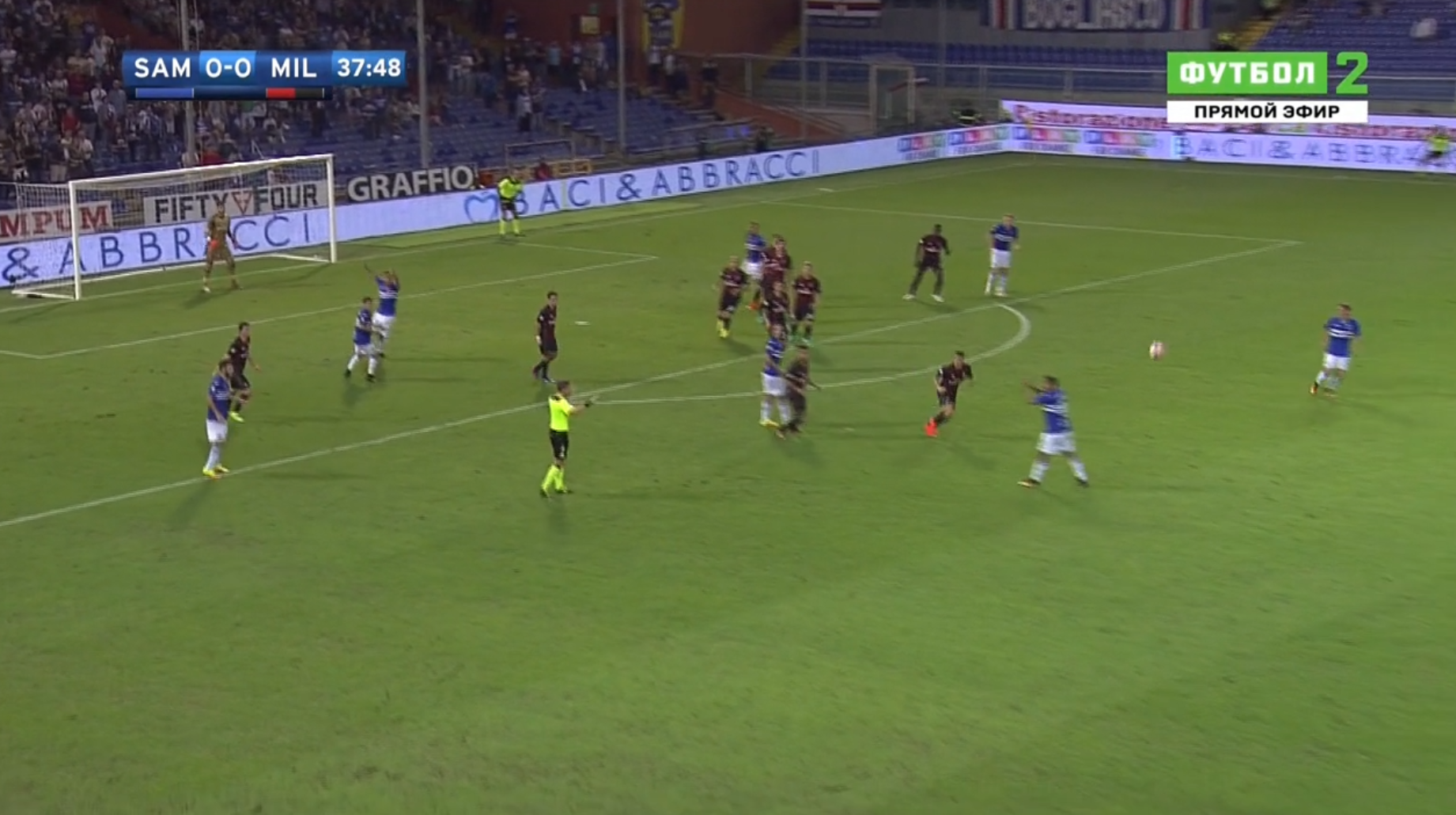}
        \caption{}
    \end{subfigure}
    \hfill
    \begin{subfigure}{0.19\textwidth}
        \includegraphics[width=\linewidth]{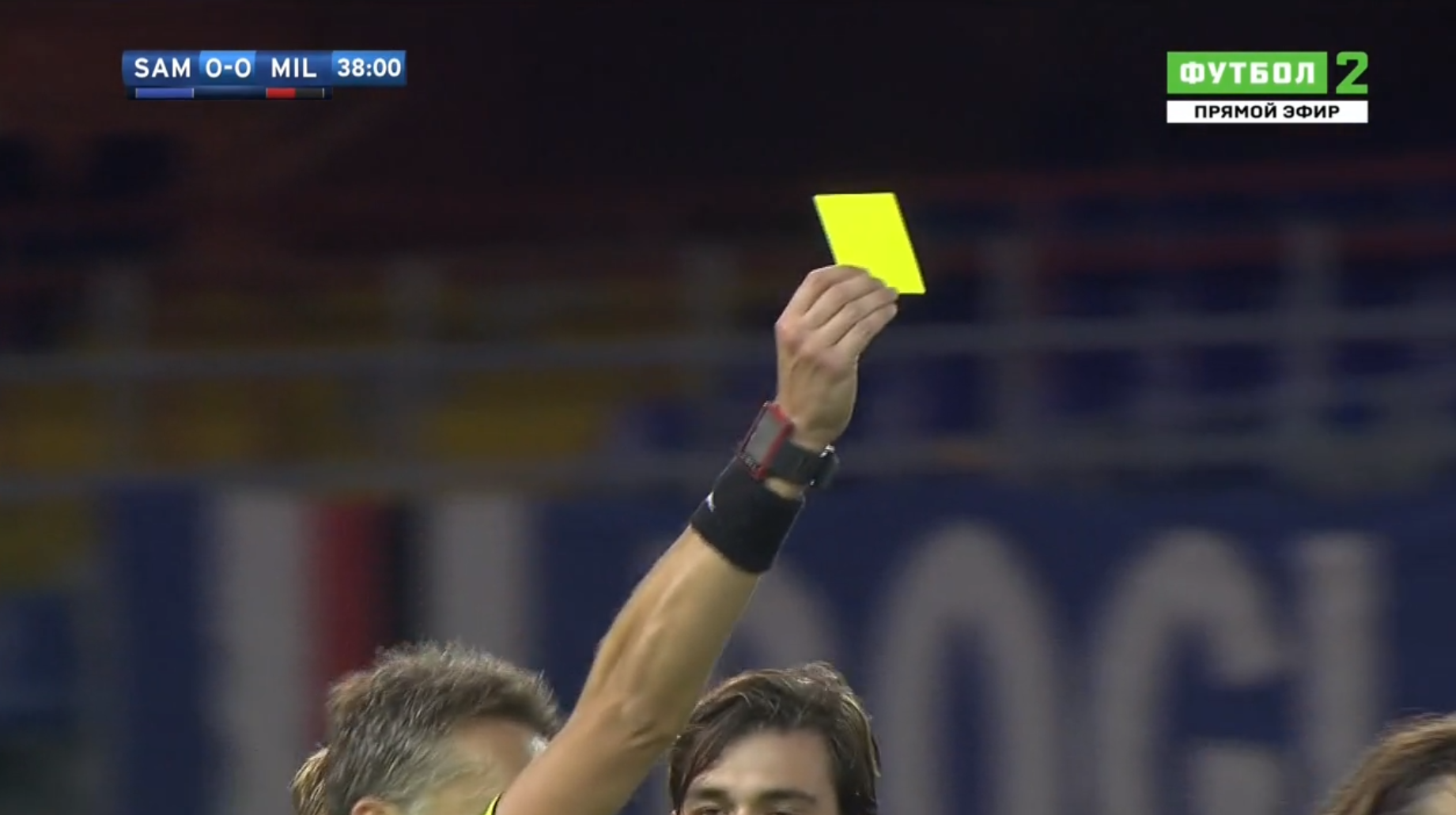}
        \caption{}
    \end{subfigure}
    \hfill
    \begin{subfigure}{0.19\textwidth}
        \includegraphics[width=\linewidth]{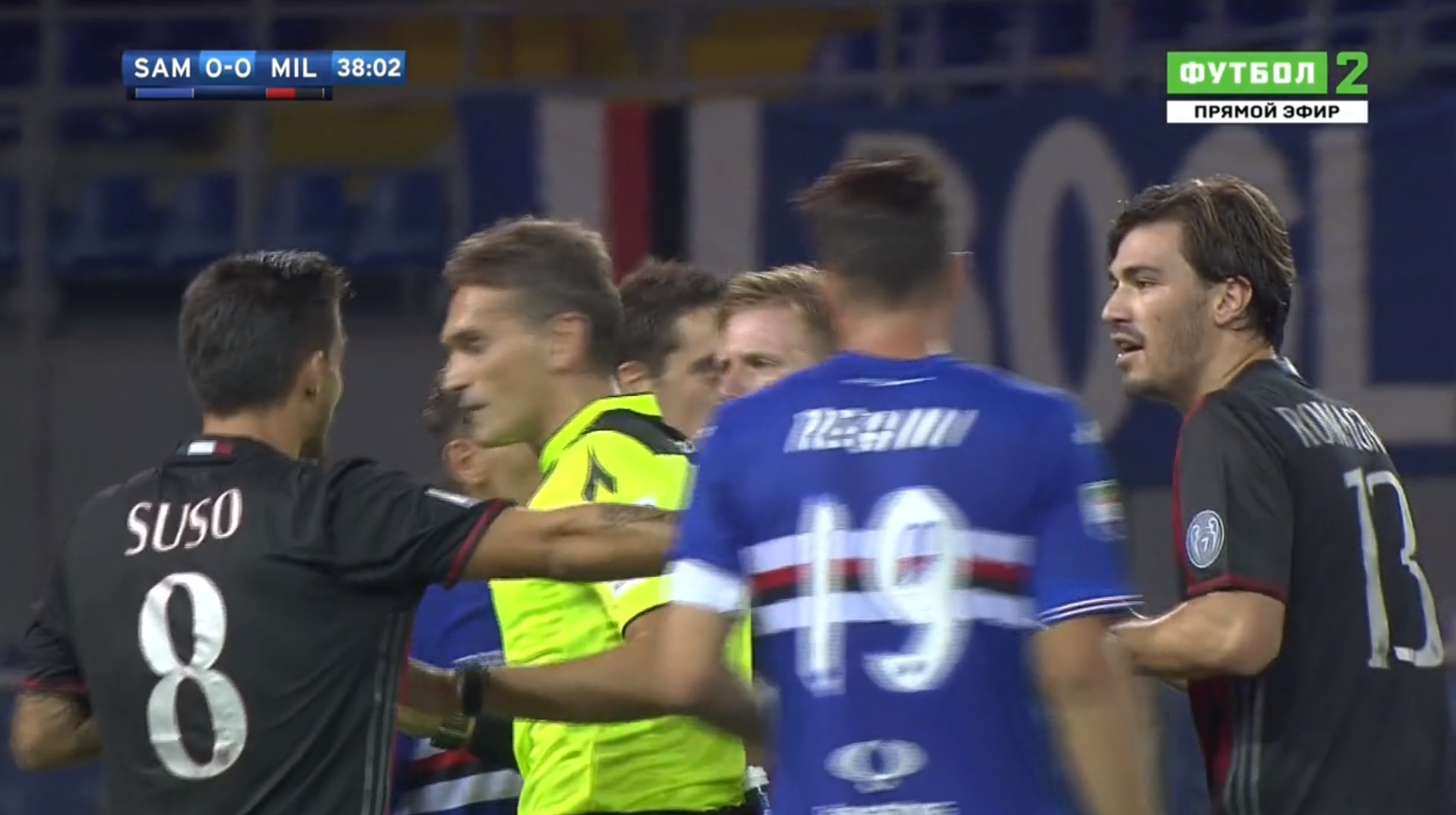}
        \caption{}
    \end{subfigure}

    \begin{tcolorbox}[colback=white,colframe=blue!75!black,boxrule=0.5mm,arc=2mm,outer arc=2mm,boxsep=2mm,left=1mm,right=1mm,top=1mm,bottom=1mm, title=Automatic Russian Transcription and English Translation]
    
        \textbf{Frame a to e (Transcription):} "\textcyrillic{Ну что, похоже будет прямой удар. Мурель попадает в Лоподулу, который, по его мнению, играл рукой. Но Лоподулу сразу подхватывает мяч. Все-таки, да, подумав, подумав, Масимилиану Арате назначает штрафной. Он послушал выгон помощника. Показывает Лоподулу желтую карточку.}"\\
        
        \textbf{Frame a to e (Translation):} "Well, it looks like it's going to be a direct hit. Murel ends up in Lopodula, who, in his opinion, played with his hand. But Lopodulu immediately picks up the ball. Still, yes, after thinking, after thinking, he appoints Masimiliana Arata penalty. He listened to the assistant's drive. Shows Lopodul a yellow card."\\

    \end{tcolorbox}
    
    \caption{Example transcription using Whisper and translation to English using Google Translate (original audio in Russian). Key frames are shown to represent the corresponding video segment.}
    \label{fig:asr-example-ru}
    
\end{figure*}

We used the language detection output from Whisper to filter non-English commentaries, and translate the non-English transcriptions into English using Google Translate~\cite{GoogleTranslate}. This choice was made due to Google Translate's accessibility, ease, and comprehensive language support. Google Translate generally provides reliable translations, particularly for common languages and everyday communication. However, its accuracy can vary depending on factors such as text complexity and linguistic nuances. Figures~\ref{fig:asr-example-de} and \ref{fig:asr-example-ru} present example transcriptions using Whisper and translations to English using Google Translate (from original game audio in German and Russian, respectively).

\subsection{Final Public Dataset}
 
The \dsname dataset and relevant code are publicly released on GitHub under \dsurl. The dataset is organized as multiple folders under the \emph{Dataset} directory, grouped by league, season, and game. Figure~\ref{fig:dataset-structure} presents an overview of the directory structure. Each \gls{asr} JSON file contains the transcribed (and translated where applicable) commentary segments for each game half, organized under the "segments" key, with each segment identified by a unique index within the respective half. Every segment records the start and end time in seconds, which delineate the temporal boundaries of the commentary, along with the transcribed speech, which provides a textual representation of the commentary content. Figure~\ref{fig:json-structure} presents the structure of the \gls{asr} JSON files. 

\section{Evaluation}\label{sec:evaluation}

For evaluating the transcription and translations in the \dsname dataset, we conducted an analysis that uses \gls{wer}, \gls{cer}, BLEU score, and word count as metrics. These metrics are crucial for discerning linguistic richness and accuracy.

\subsection{Comparison with Verified Transcriptions}

To evaluate the transcription performance, we conducted a comparative analysis using manually verified transcriptions from the GOAL dataset~\cite{GOALQi2023Oct}. This dataset comprises human-verified transcriptions from 40 game half videos across 20 games with original commentary in English from the SoccerNet dataset. We assessed the transcription performance of Whisper large-v1/v2/v3 on these 40 videos against ground truth transcriptions from the GOAL dataset using \gls{wer}, \gls{cer}, and BLEU score as metrics. The results are presented in Table~\ref{tab:goal-comparison}.

\begin{table}[h]

    \centering
    
    \begin{tabular}{|c|c|c|c|}
    
        \hline
        \textbf{Model} 
        & \textbf{WER} 
        & \textbf{CER} 
        & \textbf{BLEU} 
        \\ \hline
        Whisper large-v1 
        & 0.443 
        & 0.261 
        & 54.50 
        \\ \hline
        
        Whisper large-v2 
        & 0.458 
        & 0.269 
        & 52.59 
        \\ \hline
        
        Whisper large-v3
        & 0.551 
        & 0.341 
        & 47.97 
        \\ \hline
        
    \end{tabular}
    
    \caption{Transcription performance for different Whisper models~\cite{WhisperGitHub} against ground truth information from the GOAL dataset~\cite{GOALQi2023Oct}, averaged over the 20 games for which ground truth transcriptions are available.}
    \label{tab:goal-comparison}
    
\end{table}

\subsection{Model Selection}

We implemented a "Unique Word Count" heuristic as a criterion to identify the most effective \gls{asr} model among the different Whisper versions. This approach facilitated the selection of the best-performing model for each game half based on the diversity of vocabulary in the transcriptions, and proved particularly valuable as a heuristic for assessing model performance in scenarios prone to repeated content, often manifesting as hallucinations (erroneously repeated phrases in the transcription, which are not actually present in the audio). This metric effectively captures the ability of \gls{asr} models to generate diverse linguistic outputs, which is crucial for minimizing the impact of such repetitions. 

For each game half video, we selected the model with the highest unique word count as the "best" model for further analysis and application. This selection criterion is anchored on the premise that a higher diversity in word usage is indicative of reduced hallucinatory repetitions and, by extension, a more robust transcription performance in diverse acoustic environments. Table~\ref{tab:model-selection} presents the average unique word ratio (unique word count divided by the the number of total words) for different Whisper models, averaged over the videos for which the model was selected to be the "best". A pie chart depicting the distribution of selected Whisper models is given in Figure~\ref{fig:whisper-versions}.

\begin{table}[h]
    \centering
    \begin{tabular}{|c|c|c|c|c|}
        \hline
        \textbf{Model} 
        & \textbf{Number of Videos}
        & \textbf{Average Unique Word Ratio} 
        \\ \hline
        Whisper large-v1 
        & 316
        & 0.311 
        \\ \hline
        
        Whisper large-v2 
        & 231
        & 0.316 
        \\ \hline
        
        Whisper large-v3
        & 483
        & 0.370 
        \\ \hline

        Mixed Selection
        & 1030
        & 0.340 
        \\ \hline
        
    \end{tabular}%
    \caption{Average unique word ratio (number of unique words / number of total words) for different Whisper models. "Mixed Selection" represents the overall weighted average across a total of 1030 game half videos with valid commentary, using the "best" model for each video.}
    \label{tab:model-selection}
\end{table}
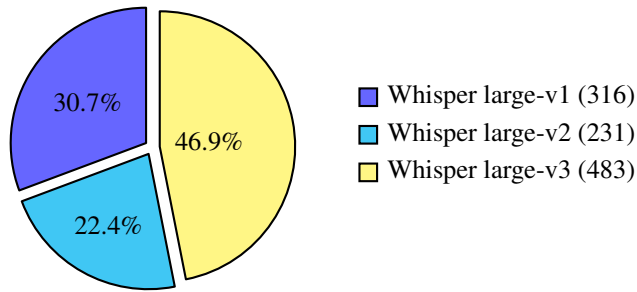
\begin{figure}[h]

    \centering
    
    \begin{tikzpicture}
        \pie[rotate=90, explode=0.1,  radius=1.8, text=legend]
        {30.7/Whisper large-v1 (316), 22.4/Whisper large-v2 (231), 46.9/Whisper large-v3 (483)}
    \end{tikzpicture}
    
    \caption{Distribution of the selected Whisper models, for a total of 1030 game half videos with valid commentary.}
    \label{fig:whisper-versions}

\end{figure}

\subsection{Vocabulary Analysis}

We conducted an in-depth examination of the linguistic patterns in the soccer game commentaries to identify prevalent verb-noun pairs, which are key to understanding the action dynamics that are described. Utilizing \gls{nlp} techniques, we parsed the transcriptions to extract named entities and specific grammatical structures, focusing on the interactions between players and teams. We then applied dependency parsing models to determine the dominant actions by identifying root verbs and their direct objects, providing clarity on the primary activities depicted.

\begin{figure*}[ht] 

    \centering
    
    \begin{subfigure}[h]{0.30\linewidth}
        \includegraphics[width=\linewidth]{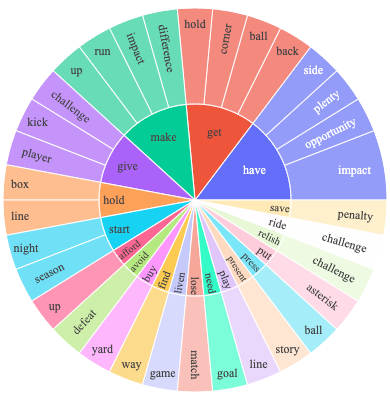}
        \caption{Ground truth transcription from GOAL dataset}
    \end{subfigure}
    \hfill
    \begin{subfigure}[h]{0.30\linewidth}
        \includegraphics[width=\linewidth]{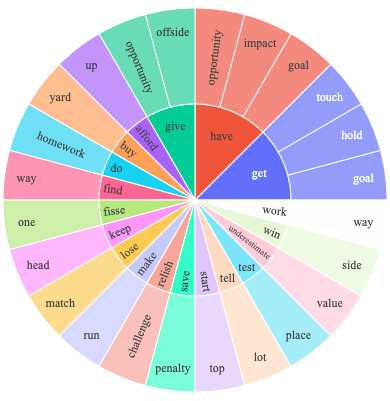}
        \caption{Whisper large-v1}
    \end{subfigure}
    \hfill
    \begin{subfigure}[h]{0.30\linewidth}
        \includegraphics[width=\linewidth]{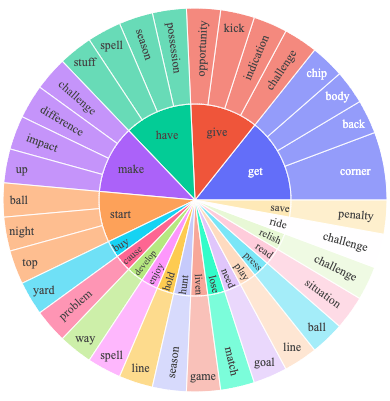}
        \caption{Whisper large-v3}
    \end{subfigure}
    
    \caption{Sunburst plots representing the frequency of verb-noun combinations in the audio commentary transcription for one game half.}
    \label{fig:vocabulary-analysis}
    
\end{figure*}

We conducted a sunburst visualization to represent the frequency and distribution of verb-noun combinations, displaying verbs as primary segments with corresponding nouns as sub-segments. This visualization, exemplified for one game half in Figure~\ref{fig:vocabulary-analysis}, not only highlights predominant actions but also showcases vocabulary diversity. Such analyses offer insights into sports broadcasting narrative styles and can guide improvements in automated commentary systems.

\section{Discussion}\label{sec:discussion}

\subsection{Dataset Applications}

The augmentation of the SoccerNet dataset to include \gls{asr} data significantly enriches the multimodal nature of the dataset, enabling a broad range of applications which can leverage the combined strengths of video, audio, and now textual data derived from spoken commentary. These applications are fundamentally enabled by \textit{enhanced content understanding}, i.e., the possibility to train automated systems which can more effectively understand, interpret, and generate soccer-related multimedia content, including an increased contextual awareness for actions, in-game events, highlights/statistics, and entities (e.g., teams, players, locations). Below, we discuss several potential applications of the \dsname dataset.

\textbf{Multi-modal event detection:} The integration of \gls{asr} data can enable more sophisticated and accurate event detection mechanisms. Combining audio cues (such as crowd noise and commentator excitement) and textual triggers from the transcription (such as specific phrases indicating goals or fouls) with visual data can improve the precision and recall of event detection. Such integration is pivotal for real-time analytics, automated highlight generation, and conducting more detailed statistical analysis. The transcription example depicted in Figure~\ref{fig:asr-example-de} showcases the advantages of \gls{asr} in this context. In scenarios where the sole analysis of video frames (e.g., based on the zoomed-out shots of the soccer pitch as shown in Figures~\ref{fig:asr-example-de-1}-\ref{fig:asr-example-de-5}) might fail to yield detailed information, \gls{asr} can provide contextual information, and allow for the extraction of additional details (such as the identification of the teams, and whether they are wearing home or away jerseys), which might not be discernible from the visual data alone. 

\textbf{Game summarization:} Automated systems can generate insightful and context-rich summaries of soccer games using the textual modality. This involves not just identifying key events, but also understanding their significance within the game's narrative, as well as the entities involved. For example, the frequency with which a player's name is mentioned during a crucial moment in the game can indicate the level of their influence, which can be used for assigning an appropriate amount of the overall word budget for the player in the description of the action. 

\textbf{Automated soccer commentator:} Commentary datasets such as \dsname up exciting possibilities for the development of automated real-time soccer game commentator systems. With increasingly more datasets available for training, models can be trained to create more realistic and contextually relevant, but also dynamic and captivating commentary, potentially offering elevated levels of immersion and engagement for the fans. 

\textbf{Tactical analysis:} In certain cases, the \gls{asr} data can also provide a context for tactical analysis, by enabling an understanding of the discussions made by commentators based on the tactical adjustments made by the coaches during soccer games. In this sense, spoken commentary provides a layer of strategic insight that is typically not captured by video and audio alone. These insights can then also be integrated with various audio cues (such as the audio intensity levels from the stadium) to pinpoint and map tactical highlights to important fan moments within the game.

These applications demonstrate the significant potential of the \dsname dataset to contribute to sports research and broadcasting. Using the power of multimodal data, researchers and developers can create more sophisticated tools that address a wide range of needs within the sports community.

\subsection{Limitations and Future Work}

Our use of deep learning models for \gls{asr} and batch translation for the curation of the \dsname dataset have also introduced a number of limitations. We aim to address several of the following challenges in future work. 

\textbf{Transcription accuracy:} The inference capability and accuracy of the \gls{asr} models are inherently constrained by their design, occasionally leading to errors in transcription. 

\textbf{Hallucinations:} Whisper models are known to be prone to hallucinations (unwarranted repetition of phrases and words). During the curation of the \dsname dataset, we have observed hallucinations across all Whisper \gls{asr} models, especially under conditions where the audio inputs were devoid of human speech, were excessively noisy, or contained musical elements. Such conditions significantly challenge transcription accuracy of the models, leading to the anomalous generation of repeated phrases. These repetitions not only degrade the quality of the transcription, but also affect the usability of the transcribed text in downstream applications, such as subtitle generation or voice-driven gameplay interaction. By analyzing the occurrences of repeated words under various audio conditions, we aimed to understand the strengths and limitations of different models. 
    
\textbf{Audio quality:} As mentioned above, we have seen that the quality of the audio to be transcribed significantly impacts the reliability of the transcriptions. Implementing advanced audio pre-processing techniques to filter out background noise and music could enhance the clarity of the input signal for \gls{asr} systems. Combining Whisper with other technologies such as \gls{vad} could also help reduce \gls{asr} errors and hallucinations.

\textbf{Human verification:} The absence of human-verified annotations to serve as ground truth poses a substantial challenge for all \gls{asr} datasets. Generating such annotations is prohibitively expensive, thus limiting the ability of researchers to comprehensively verify and refine \gls{asr} datasets. However, this limitation is also critical to address, in order to advance the reliability and applicability of our \dsname dataset. We therefore plan to invest in the curation of human-verified annotations, for a subset of the assets in \dsname, as future work. Despite being expensive, we believe that such a dataset could provide a valuable benchmark for assessing and improving \gls{asr} accuracy. Efforts could also involve collaborations with various academic institutions and community-sourcing the annotations.

\textbf{Batch translations:} 
Machine translation systems, such as Google Translate, are not infallible and their limitations must be acknowledged. When translations are performed on entire texts at once, they can produce different outputs compared to translations of specific, smaller segments. 
For instance, as depicted in Figure~\ref{fig:asr-example-ru}, the Russian term \textcyrillic{"штрафной"} was erroneously translated as "penalty kick" when the entire game half was translated, contrasting with the accurate translation "free kick" when only the segment was translated. This discrepancy underscores the contextual challenges faced by machine translation.
\section{Conclusion}\label{sec:conclusion}

The augmentation of the SoccerNet dataset using \gls{asr} technology marks a pivotal advance in the domain of sports analytics, providing a richer and more integrated approach to understanding soccer games. The inclusion of multimodal data not only aids in the accurate detection and description of in-game events but also enriches the dataset’s applicability across various analytical tasks such as sentiment analysis and tactical assessments. Despite the limitations associated with \gls{asr}, such as potential inaccuracies and hallucinations under challenging audio conditions, the benefits presented by the \dsname dataset are substantial. Future work will focus on refining \gls{asr} accuracy through the curation of a human-verified annotation subset and the implementation of advanced audio pre-processing techniques. By continually improving the dataset’s quality and the methodologies used to analyze it, \dsname is poised to significantly influence the development of automated systems for sports broadcasting and analytics, enhancing the consumption and comprehension of sports events globally. This work not only underscores the transformative impact of integrating \gls{asr} into sports video analysis but also sets a precedent for future research in the field, promising deeper insights and more effective tools for researchers, sports analytics professionals, and enthusiasts.

\section*{Acknowledgement}
This work was partly funded by the Research Council of Norway, project number 346671 (AI-Storyteller), and has benefited from the Experimental Infrastructure for Exploration of Exascale Computing (eX3), which is financially supported by the Research Council of Norway under contract 270053.

\interlinepenalty 10000
\bibliographystyle{plainnat}
\bibliography{abbreviation-short,references}
\clearpage
\appendix
\setcounter{secnumdepth}{0}
\section{APPENDIX}

Figure~\ref{fig:dataset-structure} presents an overview of the \dsname dataset directory under \dsurl. 

\begin{figure}[htbp]

    \begin{flushleft}
        \dirtree{%
            .1 \faDatabase\space Dataset.
            .2 \faFolder\space whisper\_v1.
            .3 \faTrophy\space england\_epl.
            .4 \faCalendar\space 2014-2015.
            .5 \faSoccerBallO\space 2016-03-02-23-00 Liverpool 3-0 Manchester City.
            .6 \faCloud\space 1\_asr.json.
            .6 \faCloud\space 2\_asr.json.
            .5 {...} .
            .4 \faCalendar\space 2015-2016.
            .4 {...} .
            .3 \faTrophy\space europe\_uefa-champions-league.
            .4 {...} .
            .3 \faTrophy\space france\_ligue-1.
            .4 {...} .
            .3 \faTrophy\space germany\_bundesliga.
            .4 {...} .
            .3 \faTrophy\space italy\_serie-a.
            .4 {...} .
            .3 \faTrophy\space spain\_laliga.
            .4 {...} .
            .2 \faFolder\space whisper\_v1\_en.
            .3 {...} .
            .2 \faFolder\space whisper\_v2.
            .3 {...} .
            .2 \faFolder\space whisper\_v2\_en.
            .3 {...} .
            .2 \faFolder\space whisper\_v3.
            .3 {...} .
            .2 \faFolder\space whisper\_v3\_en.
            .3 {...} .
        }
    \end{flushleft}
    
    \caption{\dsname dataset directory structure.} 
    \label{fig:dataset-structure}
\end{figure}

Figure~\ref{fig:json-structure} presents the structure of the \gls{asr} JSON files in the dataset.

\begin{figure}[h]

    \begin{lstlisting}]
    {
      "segments": {
        (int) <segment index>:[
            (float) <start time in seconds>,
            (float) <end time in seconds>,
            (string) <commentary text>
        ],
        ....
      }
    }
    \end{lstlisting}
    
    \caption{Structure of the \gls{asr} JSON files in the \dsname dataset.}
    \label{fig:json-structure}

\end{figure}

\end{document}